# Exploring and Visualizing COVID-19 Trends in India: Vulnerabilities and Mitigation Strategies

1st Swayamjit Saha
*Dept. of Computer Science and Engineering*
*Mississippi State University*
665 George Perry St, Mississippi State, MS 39762, USA
ss4706@msstate.edu

2nd Kuntal Ghosh
*Center for Soft Computing Research*
*Indian Statistical Institute*
203, BT Road, Kolkata, 700108, West Bengal, India
kuntal@isical.ac.in

3rd Garga Chatterjee
*Psychology Research Unit*
*Indian Statistical Institute*
203, BT Road, Kolkata, 700108, West Bengal, India
garga@isical.ac.in

4th J. Edward Swan II
*Dept. of Computer Science and Engineering*
*Mississippi State University*
665 George Perry St, Mississippi State, MS 39762, USA
swan@cse.msstate.edu

*Abstract*—Visualizing data plays a pivotal role in portraying important scientific information. Hence, visualization techniques aid in displaying relevant graphical interpretations from the varied structures of data, which is found otherwise. In this paper, we explore the COVID-19 pandemic influence trends in the subcontinent of India in the context of how far the infection rate spiked in the year 2020 and how the public health division of the country India has helped to curb the spread of the novel virus by installing vaccination centers across the diaspora of the country. The paper contributes to the empirical study of understanding the impact caused by the novel virus to the country by doing extensive explanatory data analysis of the data collected from the official government portal. Our work contributes to the understanding that data visualization is prime in understanding public health problems and beyond and taking necessary measures to curb the existing pandemic.

## I. INTRODUCTION

The emergence of severe acute respiratory syndrome coronavirus 2 (SARS-CoV-2) in late 2019 marked the beginning of a global health crisis unprecedented in recent history. Originating in Wuhan, China, initially linked to the Huanan Seafood Wholesale Market [4], this novel coronavirus quickly spread all across the world, leading the World Health Organization (WHO) to declare a pandemic to the world by March 2020. SARS-CoV-2, a member of the betacoronavirus genus Sarbecovirus [5], shares genetic similarities with the virus responsible for the 2002-2003 SARS outbreak. The same virus transmitted rapidly among humans and posed significant mortality rate thereby imposing immense challenges to the global health systems, economies, and societal norms [2].

Data is intuitive without any doubt. We tend to live in a data-driven world, where the influence of data is significant to a maximum extent. Trillions of Gigabytes of data are produced by our planetary people in a daily basis. An individual is responsible for producing about 1.7 megabytes of data every second in the planet [1]. It is evident why organisations try to capture the user data from the various platforms instilled by them. As a consequence of the ever-increasing growth of the information systems, the data is collected in them. However, fewer techniques have been incorporated to examine and understand the data. As opposed to the monitoring of the data, the paradigm has shifted pragmatically to interpret and analyse the same data. In other words, it is critical to examine and understand the data that we have so that we can delve deep into the insights of the data and fetch meaningful information and conclusions from the same data.

Without the support of data, there is a higher probability that people will tend to make unfair judgements based purely out of their prejudices, conventional beliefs and instincts [1]. As a consequence, they might act inaccurately based on false assumptions, beliefs or the prejudices. Hence, data acts as a key role in driving the individuals towards intelligent decisions to problems and evaluating performances of stances – something which the contemporary technologies highly rely on. Albeit, data is important, it is crucial to form a practical approach so that the examination and evaluation of the learning from the data can be obtained. The solution to such problem is data visualization.

COVID-19 pandemic has undoubtedly brought a significant amount of socio-economic, economic, and social impact to the society we live everyday. In our experiment, we tend to use data science tools for COVID-19 visualization. We visualize the initial pattern of infection spread across Kerala state in India, and how the government of India has curbed the spread of the novel virus by installing vaccine centers all across the country and incorporated vaccine dosages to the people. We discuss Related Work in Section II, the Notations and Problem Statement in Section III, the Proposed Approach in Section IV, Results and Description in Section V followed by the Conclusion.

## II. RELATED WORK

### A. Influence of Big Data for advances in technology

In the current era dominated by big data, the rapid increase in the produce of vast and diverse datasets has revolutionized numerous fields, thereby offering numerous opportunities as well as challenges. "Without data, you're just another person with an opinion" [3]. With the exponential growth in data volumes sourced from a myriad of domains such as social networks [6], financial markets [7], and health records [8], the landscape of information has expanded widely. These datasets vary in terms of accuracy and complexity, enclosing precise data as well as uncertain and imprecise data. Leveraging such big data necessitates sophisticated analytical approaches rooted in data science. Data science employs a spectrum of methodologies including data mining, machine learning, statistical modeling, and analytics to extract valuable insights and knowledge from these expansive datasets [9]. This wealth of discovered knowledge not only enhances our understanding of intricate phenomena like disease dynamics [10] but also equips researchers and policymakers with actionable insights to devise effective strategies for disease detection, prevention, and management [11].

During the COVID-19 pandemic, the significance of epidemiological data was pivotal to the researchers for striving to comprehend and combat the viral outbreak [3]. While it was evident that the medical and clinical aspects of COVID-19 has collected substantial attention of the medical researchers, computer scientists have focused on the epidemiological dimensions of the disease. Datasets collected extensively from medical fields comprising of infection rates, transmission patterns, and demographic variables were analysed, and computational approaches were framed to contribute for the crucial insights of the dynamics of the viral spread. These insights served as a direct guide to in managing public health interventions, shaping policy decisions, and preparing varied responses to mitigate such future outbreaks. The integration of data science tools and techniques with epidemiological research underscores a collaborative approach to tackling global health challenges, highlighting the transformative potential of big data in addressing emergent health crises like COVID-19.

In the realm of big data analysis, visualization serves as a powerful tool to distill complex information into easily understandable representations. More significantly in the context of COVID-19 epidemiological data, visualizations play a crucial role in conveying spatial and temporal trends of the pandemic [3]. Existing visualizers [18]–[21] predominantly focus on illustrating metrics such as confirmed cases and mortality rates across geographical regions and over time. This spatial-temporal approach enables the comparison of public health strategies and interventions, highlighting their impact on flattening epidemic curves. Moreover, integrating demographic factors such as population size and testing rates allows for normalized metrics like cases and mortality per thousand or million inhabitants, providing a more nuanced understanding of disease dynamics across different populations.

Beyond the basic metrics, the richness of epidemiological data presents opportunities for deeper insights through data mining techniques. For example, frequent pattern mining can uncover hidden relationships among attributes associated with COVID-19 cases, revealing patterns that may not be immediately apparent [12]. Visual representations of these mined patterns enhance comprehension by presenting complex relationships in an intuitive manner. Such visual analytics not only facilitate better interpretation of the underlying data but also aid in identifying critical factors influencing disease transmission and outcomes.

### B. COVID-19 Visualizer dashboards

In [13], the authors analyse a large body of research literature using both topic and visual information to distill key themes and insights. Their research analysis reveals: (a) the emergence of research on social distancing after a 70-year hiatus; (b) insights into cross-disciplinary efforts to comprehend the impacts of this unique situation; (c) trends in medical research topics; and (d) the progression of the pandemic as reflected in academic publications. The authors aimed for these methods and findings to serve as a reference for similar research systems, to inspire new research avenues, and to contribute to the ongoing battle against the pandemic. In the study [14], the authors introduced a platform designed for the semantic visualization of various entities and relationships derived from the CORD-19 dataset. Their approach involved indexing the results from multiple NLP pipelines to generate semantically labeled elements for tag clouds and heat maps. A novel technique presented in the paper is the use of parameter reduction operations on the extracted relationships, which creates "relation containers" or functional entities. These entities could be visualized using the same methods, facilitating the depiction of multiple relationships as well as both partial and complete protein-protein interaction pathways.

In [15], the authors introduce VIDAR-19, a project designed to automatically extract information about diseases and potential risk factors from the CORD-19 dataset. The project utilizes the ICD-11 disease classification system maintained by the WHO both as a data source for the extraction process and as a repository for the results. Originally developed for COVID-19, VIDAR-19 has the potential to identify risk-associated diseases and compute relevant indicators from the data. The authors display the results on a dashboard that allows users to visually explore diseases at risk within the classification hierarchy. The authors in [?] claim that VIDAR-19 has broader applicability apart from it's application in COVID-19 and can be extended to be used with datasets pertaining to other health conditions.

The C2SMART researchers in [16] present an interactive data dashboard for analyzing COVID-19 data infection patterns. The dashboard integrates data from a variety of sources, including traditional traffic sensors, crowdsourcing apps, probe vehicles, real-time traffic cameras, and police and hotline reports from cities like New York City (NYC) and Seattle in Washington. Utilizing cloud computing and data mining techniques for data acquisition and processing, the platform supports interactive visualization and analytics of various mobility and sociability metrics. The dashboard functions as a comprehensive data fusion tool for disparate open data sources, which are typically difficult to consolidate in one location, and is highly scalable, making it adaptable for use in other cities.

The authors in [17], developed a system for analyzing and visualizing the COVID-19 pandemic. Through the system, the authors have issued daily reports that collectively have garnered over 20 million page views. Public feedback indicates that these reports have significantly contributed to a clearer understanding of the pandemic, helped stabilize public sentiment, and encouraged greater cooperation with government efforts in combating COVID-19.

Researchers at WHO developed a dashboard for COVID-19 infection visualizations all across the world [18]. Similar efforts has been made by Center for Systems Science and Engineering (CSSE) [19] at Johns Hopkins University (JHU), European Center for Disease Prevention and Control (ECDC) [20], and Covid 19 Dashboard Portal developed by Ministry of Health and Family Welfare, Government of India [21].

III. NOTATIONS AND PROBLEM FORMULATION

In the realm of graphics and data visualization, the grammar of graphics offers a comprehensive framework for constructing visual representations based on data attributes. Chapters seven and eight of ggplot emphasize the fundamental concepts of geometry and aesthetics, which form the core of its implementation. While ggplot's interpretation of transformations combines variable, scale, and coordinate transformations into a unified system, this integration sacrifices some flexibility in exchange for enhanced programming simplicity. This streamlined approach facilitates easier implementation of visualizations but may limit the customization options available compared to the broader grammar of graphics framework.

Despite its streamlined approach, ggplot remains a versatile tool particularly suited for scenarios where data control may be less stringent. Unlike stricter implementations of the grammar of graphics, ggplot allows for seamless integration of data from multiple sources without explicit declarations of data linkage. This flexibility proves advantageous in practical applications, facilitating the creation of static graphics where precise data relationships may not be essential. Leveraging the extensive capabilities of R, ggplot empowers users with a rich environment for data manipulation and statistical analysis, underscoring its utility as a tool integrated within a larger computational framework rather than as a standalone application. Thus, while ggplot may sacrifice some of the comprehensive rigor of the grammar of graphics, its adaptability and integration capabilities make it a pragmatic choice for diverse data visualization needs within the R environment [22].

We extensively used the ggplot package in R programming language to draw the visual representations of the considered cases. The ggplot package use grid low-level functions to draw their plots [23].

The steps involved in the incorporation of graphics to the COVID-19 data are: I. Installing and initializing libraries like 'ggplot2', 'tidyverse' and 'maps'. II. Importing data from CSV file using read.csv() function. III. Exploring the structure of the raw data using str() function. IV. Tidying Covid 19 data: handling missing values when necessary, pivoting data whenever needed for better analysis.

The following problems were formulated and visualized accordingly. The proposed approach to the problems and the corresponding results are discussed in Sections IV and V respectively.

Problem Statement I. : Time series of Initial Cases in Kerala from 1/30/2020 to 3/1/2020

Problem Statement II. : Visualizing how the vaccine sites accommodation has changed over the time with respect to the sessions of vaccine doses

Problem Statement III. : Visualizing how the state of Odisha in India has made an exponential growth in installation of COVID-19 vaccination centers.

IV. PROPOSED APPROACH

In this section, we discuss the dataset description and the proposed approaches in R programming language to visualize the solutions to the corresponding problem statements discussed in Section III.

Primarily, we used 2 CSV files for visualizing the data from the COVID-19 in India dataset that was uploaded in Kaggle webpage [24] by researchers of ISI Bangalore viz. "Covid_19_india.csv" and "covid_vaccine_statewise.csv". State level data comes from the Ministry of Health and Family Welfare [25] and the testing data and vaccination data comes from covid19india webpage [26].

Fig. 1. Snapshot of "Covid_19_india.csv" showing the various column attributes

Fig. 2. Snapshot of "covid_vaccine_statewise.csv" showing the various column attributes

Before creating visualizations of the solutions to the above problem statements it is critical to do the following tasks: Install the necessary libraries viz. ggplot2 and tidyverse

```
R> install.packages("ggplot2")
R> install.packages("tidyverse")
```
Load the corresponding libraries
```
R> library(tidyverse)
R> library(ggplot2)
```
Importing data from CSV file
```
R> covid_data<-
   read.csv( covid _19_india.csv )
```
Explore the structure of the raw data
```
R> str(covid_data)
```
Tidying Covid 19 data
```
R>tidy_covid_data <- covid_data %>%
```
When working with data, handling missing values is crucial for accurate analysis. For instance, one approach is to drop rows containing any missing values using drop_na() function. Additionally, transforming Date columns into Date objects ensures consistency and facilitates temporal analysis, achieved here by using mutate(Date = as.Date(Date, format = "%d/%m/%Y")). Furthermore, pivoting data into long format can enhance analytical capabilities, especially when dealing with separate columns like Confirmed, Recovered, and Deaths; this transformation is achieved using pivot_longer(cols = c(Confirmed, Deaths), names_to = "Status", values_to = "Count"). These steps collectively contribute to ensuring data integrity and optimizing data structures for deeper insights and analysis.

Explore the structure of the tidy data
```
R> str(tidy_covid_data)
```
Save the tidy data to a new CSV file
```
R>write.csv(tidy_covid_data,
   tidy_covid_data.csv ,
   row.names = FALSE)
```

A. Approach to Problem Statement I We filter data for the specified date range 1/30/2020 to 3/1/2020 for Kerala state in India using:
```
R> library(ggplot2)
R> filtered_covid_data1 <- covid_data %>% fi
   lter(Sno >= 1, Sno <=32)
R> ggplot(data = filtered_covid_data1,
   mapping = aes(x = Sno, y = Confirmed))
   + geom_point(mapping =
   aes(color =    Red   ))
```

B. Approach to Problem Statement II We filter data for the specified date range 1/30/2020 to 3/1/2020 for Kerala state in India using:
```
R> library(ggplot2)
R> covid_data <-
   read.csv("covid_19_india.csv")
R> filtered_covid_data1 <-
   covid_data %>%
   filter(Sno >= 1, Sno <=32)
R> ggplot(data = filtered_covid_data1,
   mapping = aes(x = Sno, y = Confirmed))
   + geom_point(mapping = aes(color = "Red"))
```

C. Approach to Problem Statement III We filter data for the specified date range 1/30/2020 to 3/1/2020 for Kerala state in India using:
```
R> library(ggplot2)
R> covid_data4 <-
   filter(covid_data3,
   State == "Odisha")
R> ggplot(data = covid_data4,
   mapping = aes(x = Sessions, y = Sites))
   +geom_point(mapping = aes(size = Sessions),
   alpha = 1/3) + geom_smooth()
```

V. RESULTS

Our results corresponds to the graphical plots generated as a result of the solutions to the problem statements discussed in Section III.

The following Fig.3 illustrates the initial confirmed positive COVID-19 infections serially in the state of Kerala in India from the period 1/30/2020 to 3/1/2020.

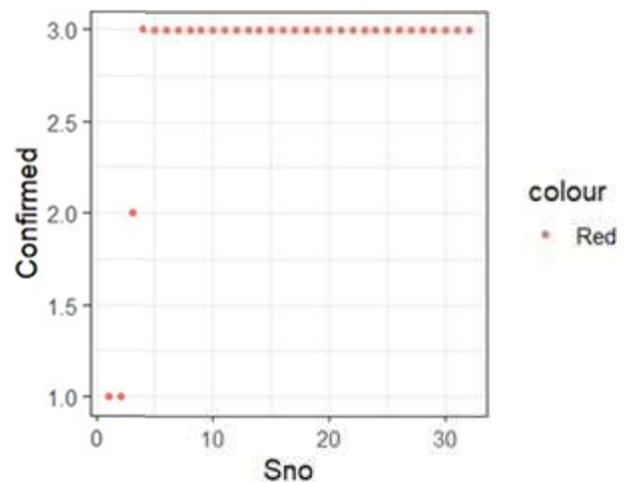

Fig. 3. Time series of Initial Cases in Kerala from 1/30/2020 to 3/1/2020.

The following Fig.4 illustrates how the Ministry of Health & Family Welfare (MoHFW) has registered COVID-19 vaccination doses in an exponentially increasing fashion among the population to curb the spread of the virus.

The following Fig.5 illustrates how the Ministry of Health & Family Welfare (MoHFW) has dramatically increased and then slowly decreased the number of vaccination centers across all states of the country with time to curb the spread of the virus among the population. The gradual decrease of vaccination centers depicts that the COVID infections gradually declined

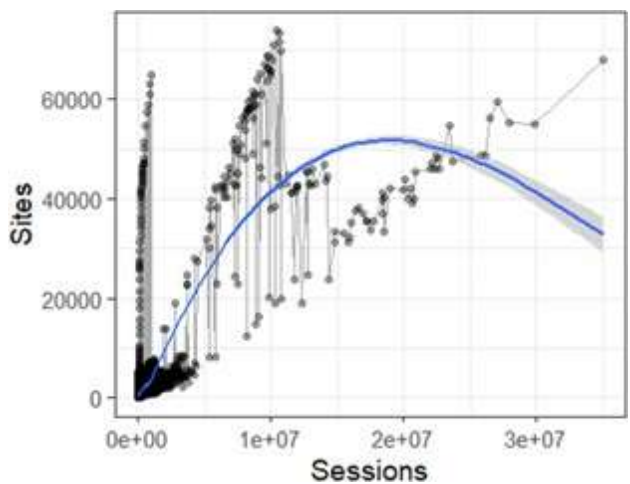

Fig. 4. Visualizing how the vaccine sites accommodation has changed over the time with respect to the sessions of vaccine doses.

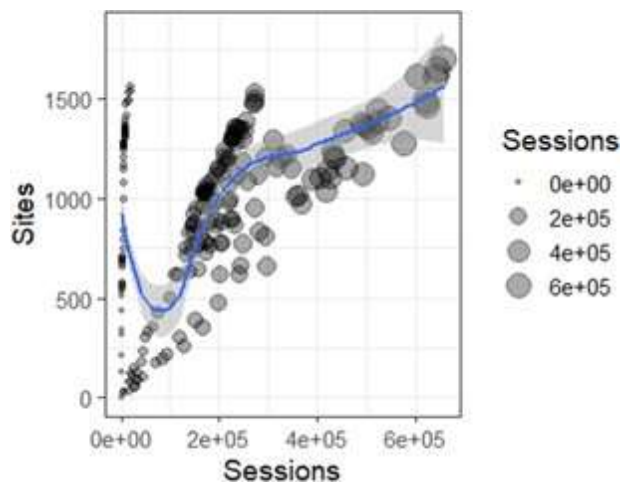

Fig. 6. Visualizing how the state of Odisha in India has made an exponential growth in installation of COVID-19 vaccination centers.

among the population promoting successful vaccination drives to control the spread of the virus.

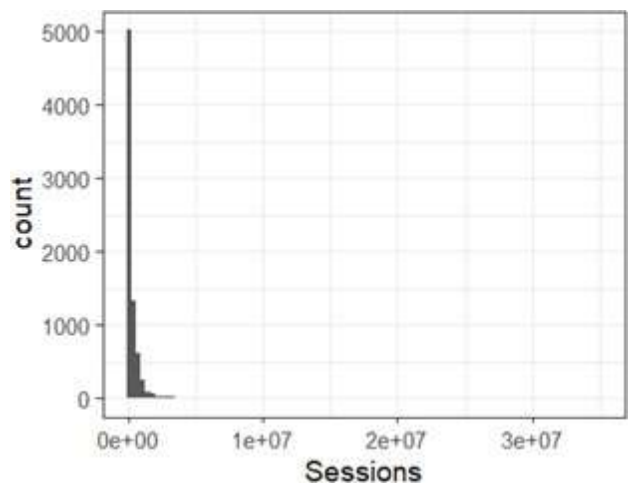

Fig. 5. Visualizing how the vaccine sites accommodation has changed over the time with respect to the sessions of vaccine doses as a histogram.

The following Fig.6 illustrates a case study of how the Ministry of Health & Family Welfare (MoHFW) has dramatically increased the number of vaccination centers exponentially for the state of Orissa to control the spread of the COVID-19 virus in the same state.

## CONCLUSION

India, faced with the formidable challenge of the COVID-19 pandemic, exhibited resilience and determination in its fight against the virus. The nation implemented a series of stringent measures, including lockdowns, testing campaigns, and vaccination drives, to curb the spread of the virus and protect its citizens. The government's proactive approach, combined with the tireless efforts of healthcare professionals and the cooperation of the public, played a crucial role in navigating through the crisis. India's vaccination campaign, one of the largest in the world, achieved significant milestones, covering a substantial portion of the population. While challenges persisted, the collective response showcased the strength and unity of the nation. The successful management of the pandemic underscored the importance of effective public health strategies, global collaboration, and the resilience of the Indian people in overcoming adversity.

## ACKNOWLEDGMENT

We would like to acknowledge the Ministry of Health & Family Welfare (MoHFW) for making the COVID-19 infection data in a state-wise level public. Additionally, we would like to thank the covid19india webpage for making the testing data and vaccination data public. Lastly, we would like to extend our thanks to the team at the Indian Statistical Institute Bangalore, with whose help we were able to get the historical data for the periods that we missed to collect and update the CSV files.

## REFERENCES

[1] Chau Vu, T. N. (2024). Exploration und Implementierung von Visualisierungstechniken für COVID-19-Daten (Doctoral dissertation, Hochschule für Angewandte Wissenschaften Hamburg).
[2] Ciotti, M., Ciccozzi, M., Terrinoni, A., Jiang, W. C., Wang, C. B., & Bernardini, S. (2020). The COVID-19 pandemic. Critical reviews in clinical laboratory sciences, 57(6), 365-388.
[3] Leung, C. K., Chen, Y., Hoi, C. S., Shang, S., Wen, Y., & Cuzzocrea, A. (2020, September). Big data visualization and visual analytics of COVID-19 data. In 2020 24th international conference information visualisation (iv) (pp. 415-420). IEEE.
[4] Worobey, M., Levy, J. I., Malpica Serrano, L., Crits-Christoph, A., Pekar, J. E., Goldstein, S. A., ... & Andersen, K. G. (2022). The Huanan Seafood Wholesale Market in Wuhan was the early epicenter of the COVID-19 pandemic. Science, 377(6609), 951-959.
[5] Pal, M., Berhanu, G., Desalegn, C., & Kandi, V. (2020). Severe acute respiratory syndrome coronavirus-2 (SARS-CoV-2): an update. Cureus, 12(3).

[6] Ianni, M., Masciari, E., & Sperlí, G. (2021). A survey of big data dimensions vs social networks analysis. Journal of Intelligent Information Systems, 57, 73-100.
[7] Shen, D., & Chen, S. H. (2018). Big data finance and financial markets. Big data in computational social science and humanities, 235-248.
[8] Kumar, S., & Singh, M. (2018). Big data analytics for healthcare industry: impact, applications, and tools. Big data mining and analytics, 2(1), 48-57.
[9] Martinez, I., Viles, E., & Olaizola, I. G. (2021). Data science methodologies: Current challenges and future approaches. Big Data Research, 24, 100183.
[10] Lopez, D., & Manogaran, G. (2016). Big data architecture for climate change and disease dynamics. In The Human Element of Big Data (pp. 315-346). Chapman and Hall/CRC.
[11] Borges do Nascimento, I. J., Marcolino, M. S., Abdulazeem, H. M., Weerasekara, I., Azzopardi-Muscat, N., Gonçalves, M. A., & Novillo-Ortiz, D. (2021). Impact of big data analytics on people's health: Overview of systematic reviews and recommendations for future studies. Journal of medical Internet research, 23(4), e27275.
[12] Muhammad, L. J., Islam, M. M., Usman, S. S., & Ayon, S. I. (2020). Predictive data mining models for novel coronavirus (COVID-19) infected patients' recovery. SN computer science, 1(4), 206.
[13] Bras, P. L., Gharavi, A., Robb, D. A., Vidal, A. F., Padilla, S., & Chantler, M. J. (2020). Visualising COVID-19 research. arXiv preprint arXiv:2005.06380.
[14] Tu, J., Verhagen, M., Cochran, B., & Pustejovsky, J. (2020). Exploration and discovery of the COVID-19 literature through semantic visualization. arXiv preprint arXiv:2007.01800.
[15] Wolinski, F. (2020). Visualization of diseases at risk in the COVID-19 Literature. arXiv preprint arXiv:2005.00848.
[16] Zuo, F., Wang, J., Gao, J., Ozbay, K., Ban, X. J., Shen, Y., ... & Iyer, S. (2020). An interactive data visualization and analytics tool to evaluate mobility and sociability trends during COVID-19. arXiv preprint arXiv:2006.14882.
[17] Zhang, S. H., Cai, Y., & Li, J. (2020). Visualization of COVID-19 spread based on spread and extinction indexes. Science China. Information Sciences, 63(6), 164102.
[18] World Health Organization. (n.d.). COVID-19 dashboard. https://data.who.int/dashboards/covid19/cases?n=c
[19] Johns Hopkins University. (n.d.). Coronavirus resource center: COVID-19 map. https://coronavirus.jhu.edu/map.html
[20] European Centre for Disease Prevention and Control. (n.d.). COVID-19 situation updates. https://www.ecdc.europa.eu/en/COVID-19/situation-updates
[21] Ministry of Health and Family Welfare. (n.d.). COVID-19 dashboard. https://covid19.mohfw.gov.in/
[22] Wickham, H. (2006). An introduction to ggplot: An implementation of the grammar of graphics in R. Statistics, 1.
[23] Murrell, P. (2009). R graphics. Wiley Interdisciplinary Reviews: Computational Statistics, 1(2), 216-220.
[24] Kaggle. (n.d.). Kaggle: Your machine learning and data science community. https://www.kaggle.com/
[25] Ministry of Health and Family Welfare. (n.d.). Ministry of Health and Family Welfare. https://mohfw.gov.in/
[26] COVID-19 India. (n.d.). COVID-19 India. https://www.covid19india.org/